\documentstyle[aps,psfig]{revtex}

\begin{document}
\draft
\title{$q$BCS - the BCS theory of $q$-deformed nucleon pairs}
\author{S. Shelly Sharma\thanks{%
Email: shelly@uel.br}}
\address{Departamento de F\'{i}sica, Universidade Estadual de Londrina, Londrina,\\
86051-970, PR, Brazil}
\author{N. K. Sharma}
\address{Departamento de Matem\'{a}tica, Universidade Estadual de Londrina, Londrina,%
\\
86040-970, PR, Brazil}
\maketitle

\begin{abstract}
We construct a coherent state of $q$-deformed zero coupled nucleon pairs
distributed in several single-particle orbits. Using a variational approach,
the set of equations of $q$BCS theory, to be solved self consistently for
occupation probabilities, gap parameter $\Delta ,$ and the chemical
potential $\lambda ,$ is obtained. Results for valence nucleons in nuclear
degenerate sdg major shell show that the strongly coupled zero angular
momentum nucleon pairs can be substituted by weakly coupled $q$-deformed
zero angular momentum nucleon pairs. A study of Sn isotopes reveals a well
defined universe of ($G,$ $q$) values, for which $q$BCS converges. While the 
$q$BCS and BCS show similar results for Gap parameter $\Delta $ in Sn
isotopes, the ground state energies are lower in $q$BCS. The pairing
correlations in $N$ nucleon system, increase with increasing $q$ (for $q$
real).
\end{abstract}


The problem of nucleon pairing has been of great interest to those
interested in solving the riddles about nuclear structure. BCS theory of
superconductivity\cite{BCS57} turned out to be a great ally in efforts to
take in to account short range interactions between nucleons. In many
nuclear structure calculations BCS\ is taken as the starting point as the
BCS wave function offers a good approximation to ground state of even-even
nuclei. Another approach to nucleon pairing problem is the seniority scheme.
In order to include the pairing correlations left out in these approximate
schemes, we studied the zero coupled nucleon pairs with $q$-deformations\cite
{she92} expressing these in terms of the generators of quantum group SU$%
_{q}(2)$. The quantum group SU$_{q}(2)$, a $q$-deformed version of Lie
algebra SU$(2)$, has been studied extensively\cite{Jimb85,Woro87,Pasq88},
and a $q$-deformed version of quantum harmonic oscillator developed \cite
{Macf89,Bied89}. The quantum group SU$_{q}(2)$ is more general than SU$(2)$
and contains the later as a special case. The underlying idea in using the
zero coupled nucleon pairs with $q$-deformations is that the commutation
relations of nucleon pair creation and destruction operators are modified by
the correlations as such are somewhat different in comparison with those
used in deriving the usual theories. The seniority scheme for $q$-deformed
nucleon pairs in a single $j$ orbit and zero seniority states for nuclei
with $q$-deformed nucleon pairs distributed over several orbits have also
been constructed in Ref. \cite{she92}. On the same lines Random Phase
Approximation equations for the pairing vibrations of nuclei have been
derived and applied to study pairing vibrations in Pb isotopes \cite{she94}.
Further a $q$-deformed version of quasi boson approximation for $0^{+}$
states in superconducting nuclei was developed. The $q$-deformed theories
reduce to the corresponding usual theories in the limit $q\rightarrow 1$.
The Nucleon pairing in a single $j$ shell has also been treated by Bonatsos
et. al \cite{Bona92,Bona94} by associating two $Q$-oscillators, one
describing the $J=0$ pairs and the other associated with $J\neq 0$ pairs. In
their formalism, $Q$-oscillators involved reduce to usual harmonic
oscillators as $Q\rightarrow 1$ and the deformation is introduced in a way
different from ours. Following the idea of building in correlations in to
the theory by using pair generators satisfying $q$-commutation relations, we
now construct the $q$-analog of BCS theory ($q$BCS) for nuclei. The single
orbit limit of $q$BCS is applied to nuclear sdg major shell with $\Omega =16$%
. In addition, the gap parameter and  ground state energies are calculated
for $^{114-124}$Sn, to elucidate the role played by $q$-deformation in these
nuclei.

\section{Zero Coupled $q$-deformed Nucleon Pairs}

The creation and destruction operators for a zero coupled nucleon pair in a
shell model orbit $j$ are

\begin{equation}
Z_{0}=-\frac{1}{\sqrt{2}}(A^{j}\times A^{j})^{0}\text{ and }\overline{Z}_{0}=%
\frac{1}{\sqrt{2}}(B^{j}\times B^{j})^{0},  \label{1}
\end{equation}
where $A_{jm}=a_{jm}^{\dagger };$ $B_{jm}=(-1)^{j+m}a_{j,-m}$. From the
anticommutation relations satisfied by the fermion creation and destruction
operators $a_{jm}^{\dagger }$ and $a_{j,-m}$, we can verify that with number
operator for fermions defined as $n_{op}^{j}=\sum_{m}a_{jm}^{\dagger }a_{jm}$
and $\Omega =(2j+1)/2$, 
\begin{equation}
\left[ Z_{0},\overline{Z_{0}}\right] =\frac{n_{op}-\Omega }{\Omega }\text{ ; 
}\left[ n_{op},Z_{0}\right] =2Z_{0}\text{ ; }\left[ n_{op},\overline{Z_{0}}%
\right] =-2\overline{Z_{0}}\text{ .}  \label{2}
\end{equation}
These operators are easily related to well known quasi-spin operators by
identifying 
\begin{equation}
S_{+}=\sqrt{\Omega }\,Z_{0}\text{ ; }S_{-}=\sqrt{\Omega }\,\overline{Z_{0}},%
\text{ and }S_{0}=\frac{(n_{op}-\Omega )}{2}.  \label{3}
\end{equation}
The quasi-spin operators $S_{+}$, $S_{-}$, and $S_{0}$ are the generators of
Lie algebra of SU(2) and satisfy the commutation relations of angular
momentum operators, that is 
\begin{equation}
\left[ S_{+},S_{-}\right] =2S_{0}\text{ , }\left[ S_{0},S_{\pm }\right] =\pm
S_{\pm }\text{ .}  \label{4}
\end{equation}

The generators of SU$_{q}(2)$ on the other hand satisfy the $q$-commutation
relations\cite{she92} 
\begin{equation}
\left[ S_{+}(q),S_{-}(q)\right] =\left\{ 2S_{0}(q)\right\} _{q}\text{ , }%
\left[ S_{0}(q),S_{\pm }(q)\right] =\pm S_{\pm }(q)\text{ ; }  \label{5}
\end{equation}
where $\{x\}_{q}=\frac{(q^{x}-q^{-x})}{(q-q^{-1})}$. Translated to $q$%
-deformed pair operators $Z_{0}(q)$ and $\overline{Z_{0}(q)}$ the new
commutation relations give 
\begin{equation}
\left[ Z_{0}(q),\overline{Z_{0}(q)}\right] =\frac{\left\{ n_{op}-\Omega
\right\} _{q}}{\Omega }\text{ ; }\left[ n_{op},Z_{0}(q)\right] =2Z_{0}(q)%
\text{ ; }\left[ n_{op},\overline{Z_{0}(q)}\right] =-2\overline{Z_{0}(q)}%
\text{ .}  \label{6}
\end{equation}

\section{The Trial Wave Function}

The proposed trial wave function for $N$ nucleons distributed over $m$
single particle orbits is, $\Psi =\Phi _{j_{1}}\Phi _{j_{2}}...\Phi _{j_{m}}$%
, where for the orbit $j,$ 
\begin{equation}
\Phi _{j}=u_{j}^{\Omega _{j}}\sum_{n=0}^{\Omega _{j}}\left( \frac{v_{j}}{%
u_{j}}\right) ^{n}\left[ \frac{\Omega _{j}!}{n!(\Omega _{j}-n)!}\right] ^{%
\frac{1}{2}}\left| n\right\rangle \text{ ;\ }\Omega _{j}=\frac{2j+1}{2}\text{
}  \label{7}
\end{equation}
and 
\[
\text{ }\left| n\right\rangle =\left[ \frac{\left\{ \Omega _{j}-n\right\}
_{q}!}{\left\{ n\right\} _{q}!\left\{ \Omega _{j}\right\} _{q}!}\right] ^{%
\frac{1}{2}}\left( S_{j+}(q)\right) ^{n}\left| 0\right\rangle \text{ } 
\]
is the normalized wave function for $n$ zero coupled nucleon pairs with $q$%
-deformation occupying single particle orbit $j$. The function $\Psi $ is
normalized in case, $u_{j}^{2}+v_{j}^{2}=1,$ for all single particle orbits.
The single particle plus pairing Hamiltonian for $q$-deformed pairs is given
by 
\begin{equation}
H=\sum_{r}\varepsilon _{r}n_{op}^{r}-G\sum\limits_{rs}S_{r+}(q)S_{s-}(q)%
\text{ };\;\text{where }r,s\equiv j_{1},j_{2},.......j_{m}\text{.}  \label{8}
\end{equation}
The matrix element $\left\langle \Psi \left| -G\,S_{r+}(q)S_{r-}(q)\right|
\Psi \right\rangle $ , obtained by using the $q$-commutation relations given
in Eq. (\ref{5}) and ignoring terms involving products of the type $%
v_{r}^{4}u_{r}^{m}(m=2,4,..,\Omega _{r})$, is found to be 
\[
\left\langle \Psi \left| -G\,S_{r+}(q)S_{r-}(q)\right| \Psi \right\rangle
=-G\,v_{r}^{2}\Omega _{r}\left\{ \Omega _{r}\right\}
_{q}+G\,v_{r}^{4}(\Omega _{r}-1)\left\{ \Omega _{r}\right\} _{q}. 
\]
We also calculate the gap parameter, 
\[
\Delta (q)=G\left\langle \Psi \left| \sum\limits_{r}S_{r+}(q)\right| \Psi
\right\rangle =\sum\limits_{r}\Delta
_{r}(q)=\sum\limits_{r}Gu_{r}v_{r}\left\{ \Omega _{r}\right\} _{q}. 
\]
Again the terms involving products of the type $v_{r}^{3}u_{r}^{m}$ have
been ignored. After these considerations, we can write the matrix element of
the Hamiltonian $H$ as 
\[
\left\langle \Psi \left| H\right| \Psi \right\rangle =\sum_{r}\left(
2\varepsilon _{r}\,\Omega _{r}\,v_{r}^{2}-G\,v_{r}^{2}\,\Omega _{r}\left\{
\Omega _{r}\right\} _{q}+G\,v_{r}^{4}(\Omega _{r}-1)\left\{ \Omega
_{r}\right\} _{q}+\frac{\Delta _{r}^{2}(q)}{G}\right) -\frac{\left( \Delta
(q)\right) ^{2}}{G} 
\]

\section{$q$BCS Gap equation and the Ground State Energy}

In order to evaluate the ground state energy of $N$ nucleons, we minimize
the expectation value of the Hamiltonian subject to the number constraint by
varying $v_{j}$ and obtain $m$ equations to be solved self consistently $,$ 
\begin{equation}
4(\varepsilon _{j}^{\prime }-\lambda )v_{j}\Omega _{j}-2\,\Delta
(q)\,\left\{ \Omega _{j}\right\} _{q}\left( \frac{1-2v_{j}^{2}}{u_{j}}%
\right) -4G\,v_{j}^{3}\left\{ \Omega _{j}\right\} _{q}\left( \left\{ \Omega
_{j}\right\} _{q}-\Omega _{j}+1\right) =0,  \label{9b}
\end{equation}
where $\varepsilon _{j}^{\prime }=\varepsilon _{j}+\frac{G\left\{ \Omega
_{j}\right\} _{q}\left( \left\{ \Omega _{j}\right\} _{q}-\Omega _{j}\right) 
}{2\Omega _{j}}$. Leaving out for the time being, the term containing $%
u_{j}v_{j}^{3},$ we solve these equations to obtain the occupancies, 
\begin{equation}
v_{j}^{2}=0.5\left( 1-\frac{\varepsilon _{j}^{\prime }-\lambda }{\sqrt{%
\left( \varepsilon _{j}^{\prime }-\lambda \right) ^{2}+\left( \Delta (q)%
\frac{\left\{ \Omega _{j}\right\} _{q}}{\Omega _{j}}\right) ^{2}}}\right) ,
\label{10}
\end{equation}
gap parameter 
\begin{equation}
\Delta (q)=\sum_{j}G\,\left\{ \Omega _{j}\right\} _{q}0.5\left( 1-\frac{%
\left( \varepsilon _{j}^{\prime }-\lambda \right) ^{2}}{\left( \varepsilon
_{j}^{\prime }-\lambda \right) ^{2}+\left( \Delta (q)\frac{\left\{ \Omega
_{j}\right\} _{q}}{\Omega _{j}}\right) ^{2}}\right) ^{\frac{1}{2}}
\label{11}
\end{equation}
and consequently the gap equation 
\begin{equation}
\frac{G}{2}\sum_{j}\frac{\left\{ \Omega _{j}\right\} _{q}^{2}}{\sqrt{\left(
\varepsilon _{j}^{\prime }-\lambda \right) ^{2}\Omega _{j}^{2}+\left( \Delta
(q)\left\{ \Omega _{j}\right\} _{q}\right) ^{2}}}=1.  \label{12}
\end{equation}
To include the effect of terms containing $u_{j}v_{j}^{3}$ left out earlier,
we now replace $\lambda $ by 
\begin{equation}
\lambda (q)=\lambda +\frac{Gv_{j}^{2}\left\{ \Omega _{j}\right\} _{q}}{%
\Omega _{j}}\left( \left\{ \Omega _{j}\right\} _{q}-\Omega _{j}+1\right) .
\label{14}
\end{equation}
The ground state BCS\ energy, $\left\langle \Psi \left| H\right| \Psi
\right\rangle $ is 
\begin{equation}
E_{bcs}(q)=\sum_{j=1}^{m}\left( 2\varepsilon _{j}^{\prime }\,\Omega
_{j}\,v_{j}^{2}-G\,v_{j}^{4}\left\{ \Omega _{j}\right\} _{q}\left( \left\{
\Omega _{j}\right\} _{q}-\Omega _{j}+1\right) \right) -\frac{\left( \Delta
(q)\right) ^{2}}{G}  \label{15}
\end{equation}
We notice that in a very natural way, the SU$_{q}$(2) symmetry introduces in
the interaction energy, a $q$ dependence which is linked to the $j$-value of
the orbit occupied by the zero coupled nucleon pairs.

\section{Single orbit with 2$\Omega $ degenerate states}

A very special situation arises, when the $N$ nucleons occupy a single orbit
with an occupancy of $2\Omega $. Using the results of the previous section,
the ground state wave function is now $\Psi =\Phi _{j}$ and the ground state
energy $E_{bcs}(q)$ is 
\begin{equation}
E_{bcs}(q)=\varepsilon _{j}N-G\left\{ \Omega _{j}\right\} _{q}\frac{N}{%
4\Omega }\left( 2\left\{ \Omega _{j}\right\} _{q}-N+\frac{N}{\Omega }\right) 
\label{18}
\end{equation}
to be compared with the exact energy of the $N$ nucleon zero seniority state%
\cite{law}, 
\begin{equation}
E_{exact}=\varepsilon _{j}N-G^{\prime }\frac{N}{4}\left( 2\Omega
_{j}-N+2\right)   \label{19}
\end{equation}
We notice that we can have $E_{bcs}(q)=E_{exact}$ by choosing $q$ value and
pairing strength $G$ such that 
\[
G=\frac{G^{\prime }\Omega _{j}\left( 2\Omega _{j}-N+2\right) }{\left\{
\Omega _{j}\right\} _{q}\left( 2\left\{ \Omega _{j}\right\} _{q}-N+\frac{N}{%
\Omega }\right) }
\]
for the choice $\varepsilon _{j}=0.0$. For the special case of nuclear sdg
major shell with $\Omega =16$ , and $4,14,20,30$ valence nucleons occupying
degenerate $1d_{\frac{5}{2}},0g_{\frac{7}{2}},2s_{\frac{1}{2}},1d_{\frac{3}{2%
}},$ and $0h_{\frac{11}{2}}$ orbits, we plot $G$ versus $q$ in Fig.1 such
that $E_{bcs}(q)=E_{exact}(G^{\prime }=0.187$ MeV, $\varepsilon _{j}=0.0$
for all levels$)$. The intensity of pairing strength required to reproduce $%
E_{exact}$ is seen to fall with increasing $q$ and ultimately $G\rightarrow 0
$ for all cases$.$ From the plot at hand we can say that strongly coupled
zero coupled pairs of BCS\ theory may well be replaced by weakly coupled $q$%
-deformed zero coupled pairs of $q$BCS theory. The natural question is, is
it possible to replace the pairing interaction by a suitable commutation
relation between the pairs determined by a characteristic $q$ value for the
system at hand? To get some clues to the answer, we next consider real
nuclei for which we can get the pairing gap from the experiments.

\section{Sn Isotopes}

We examine the heavy Sn isotopes with $N=14,16,18,20,22,$ and $24$ neutrons
outside $_{50}^{100}$Sn$_{50}$ core. The model space includes $1d_{\frac{5}{2%
}},0g_{\frac{7}{2}},2s_{\frac{1}{2}},1d_{\frac{3}{2}},$ and $0h_{\frac{11}{2}%
}$, single particle orbits, with excitation energies $0.0$, $0.22$, $1.90$, $%
2.20$, and $2.80$ MeV respectively. Fig. 2 is a plot of pairing correlations
function $D=\Delta (q)/\sqrt{G}$ versus $G$ for $N=20$ in the cases where
deformation parameter takes some typical successively increasing values
varying from $1.0$ to $1.7$. We notice that in $_{50}^{120}$Sn$_{70}$,
pairing correlations increase as $q$ increases if the pairing strength $G$
is kept fixed. For $q=1.0$ that is conventional BCS theory the pairing
correlation vanishes for $G<G_{c}(\sim 0.065$ MeV$)$ as expected. As the
deformation $q$ of zero coupled pairs increases we find $D$ going to zero
for successively lower values of coupling strength, for example $G_{c}\sim
0.04$ MeV for $q=1.3$ . We may infer that the $q$BCS takes us beyond BCS
theory.

The sets of $G,q$ values that reproduce the empirical $\Delta $ for $%
_{50}^{120}$Sn$_{70}$,  are used to calculate the gap parameter $\Delta $
and the ground state BCS energy E$_{N},$ for even isotopes $^{114-124}$Sn
displayed in Fig. 3. The experimental values of $\Delta $ (filled triangles
up) are also shown. As far as the gap parameter $\Delta $ is concerned all
the sets of $G,q$ values fair equally in comparison with the experiment. The
ground state energies from $q$BCS are however in general lower than those
calculated by using BCS. The underlying $q$-deformed nucleon pairs show
increasingly strong binding as the value of $q$ is increased. It opens the
possibility of obtaining the exact correlation energies by choosing
appropriately the combination of $G,q$ values.

\section{Conclusions}

By looking at the results for $4,14,20,30$ valence nucleons in nuclear
degenerate sdg major shell, we find that the strongly coupled zero angular
momentum nucleon pairs may be replaced by weakly coupled $q$-deformed zero
angular momentum nucleon pairs. The study of a realistic case i.e. Sn
isotopes also indicates that their is a well defined universe of sets of
values for pairing strength $G$ and deformation parameter $q$, for which $q$%
BCS converges and has a non-trivial solution. For $_{50}^{120}$Sn$_{70}$ we
observe that by choosing the pairing strength $G\leq 0.217$ MeV a matching
value of deformation parameter $q$ can be found such that the experimental
pairing gap is reproduced. For the choice $G=0.07$ MeV for example a large
deformation of  $q=1.7$ is needed to reproduce the empirical $\Delta $ for $%
_{50}^{120}$Sn$_{70}$. The results of $q$BCS for Sn isotopes are not much
different from BCS as far as the Gap parameter $\Delta $ is concerned. The
ground state binding energies are however lowered by the deformation. The
pairing correlations, measured by $D=\Delta (q)/\sqrt{G}$, are seen to
increase as $q$ increases (for $q$ real) while the pairing strength $G$ is
kept fixed, in Sn isotopes. It is immediately seen that $q$ parameter is a
very good measure of the pairing correlations left out in the conventional
BCS theory.

The results of our present study are consistent with our earlier conclusions 
\cite{she92,she94} that the $q$-deformed pairs with $q>1$ $(q$ real) are
more strongly bound than the pairs with zero deformation and the binding
energy increases with increase in the value of parameter $q$. In contrast by
using complex $q$ values one can construct zero coupled deformed pairs with
lower binding energy in comparison with the no deformation zero coupled
nucleon pairs\cite{she94}. In general the pairing correlations in $N$
nucleon system, measured by $D=\Delta (q)/\sqrt{G}$ , increase with
increasing $q$ (for $q$ real) and $q$BCS takes us beyond the BCS theory. The
formalism can be tested for several other systems, for example metal grains,
where cooper pairing plays an important role.

{\LARGE Acknowledgments}

S. Shelly Sharma acknowledges support from Universidade Estadual de Londrina.

\begin{figure}
\psfig{figure=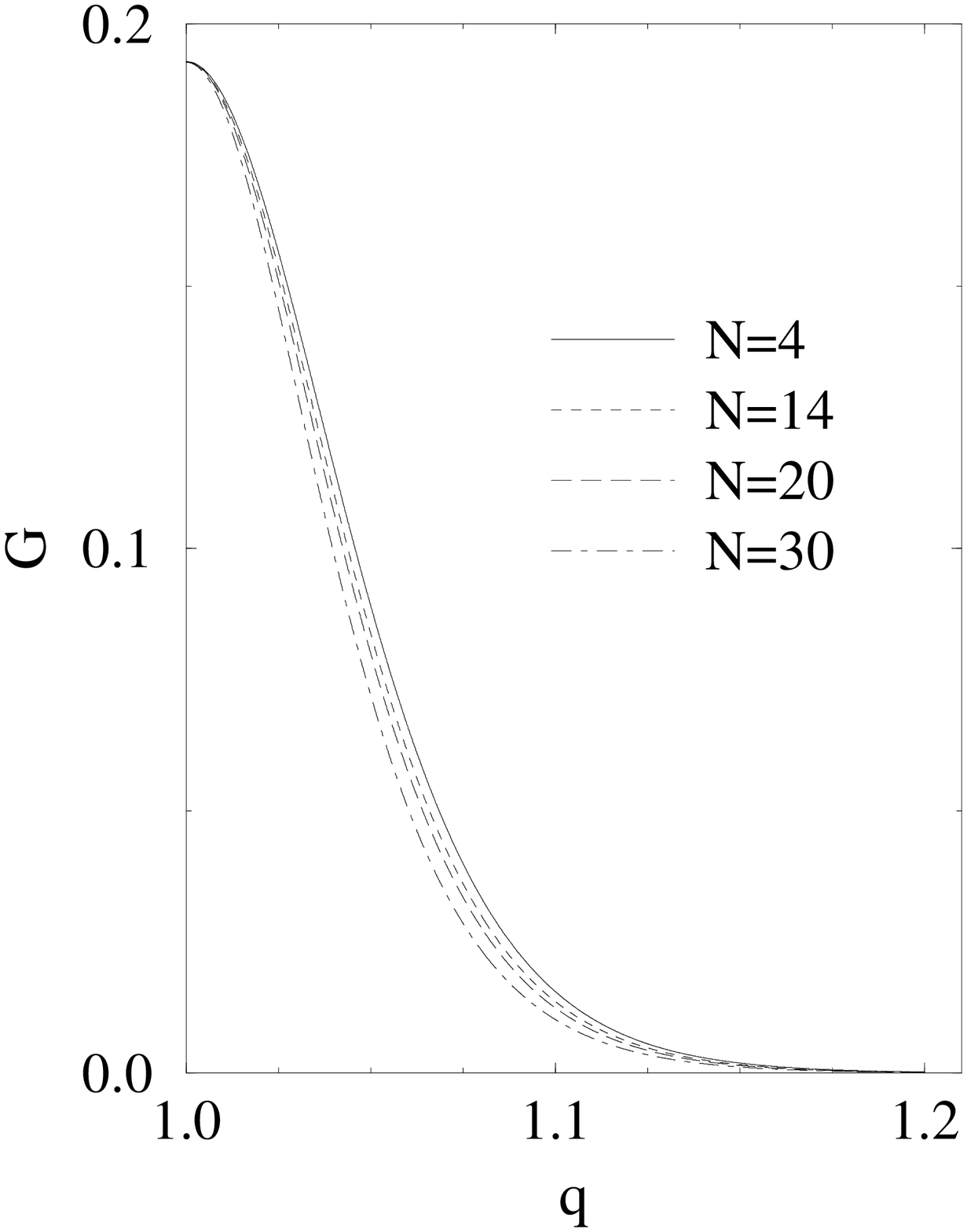,width=6in,height=8in}
\caption{$G$ versus $q$ for $4,14,20,30$ valence nucleons in sdg major
shell with $\Omega =16$ such that $E_{bcs}(q)=E_{exact},$ $G^{\prime }=0.187$
MeV, ($\varepsilon _{j}=0.0$ for all single-particle orbits)$.$}
\label{fig1}
\end{figure}

\begin{figure}
\psfig{figure=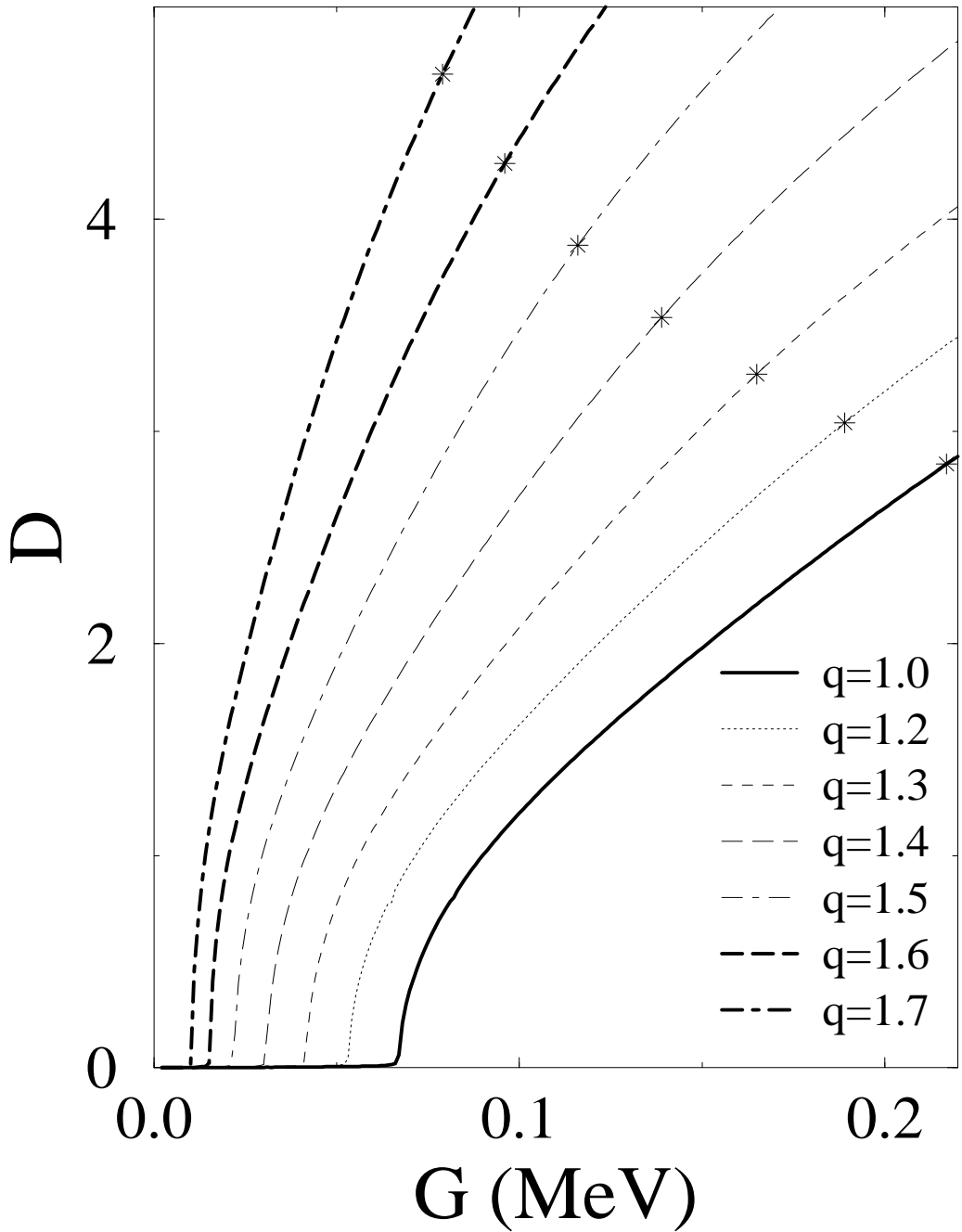,width=6in,height=8in}
\caption{The calculated pairing correlations function $D$ versus $G$ for $%
N=20$ and deformation parameter values $q=1.0,1.2,1.3,1.4,1.5,1.6 and 1.7$.
Stars on the curves mark the G value that reproduces empirical $\Delta $
for $ ^{120}$Sn  }
\label{fig2}
\end{figure}

\begin{figure}
\psfig{figure=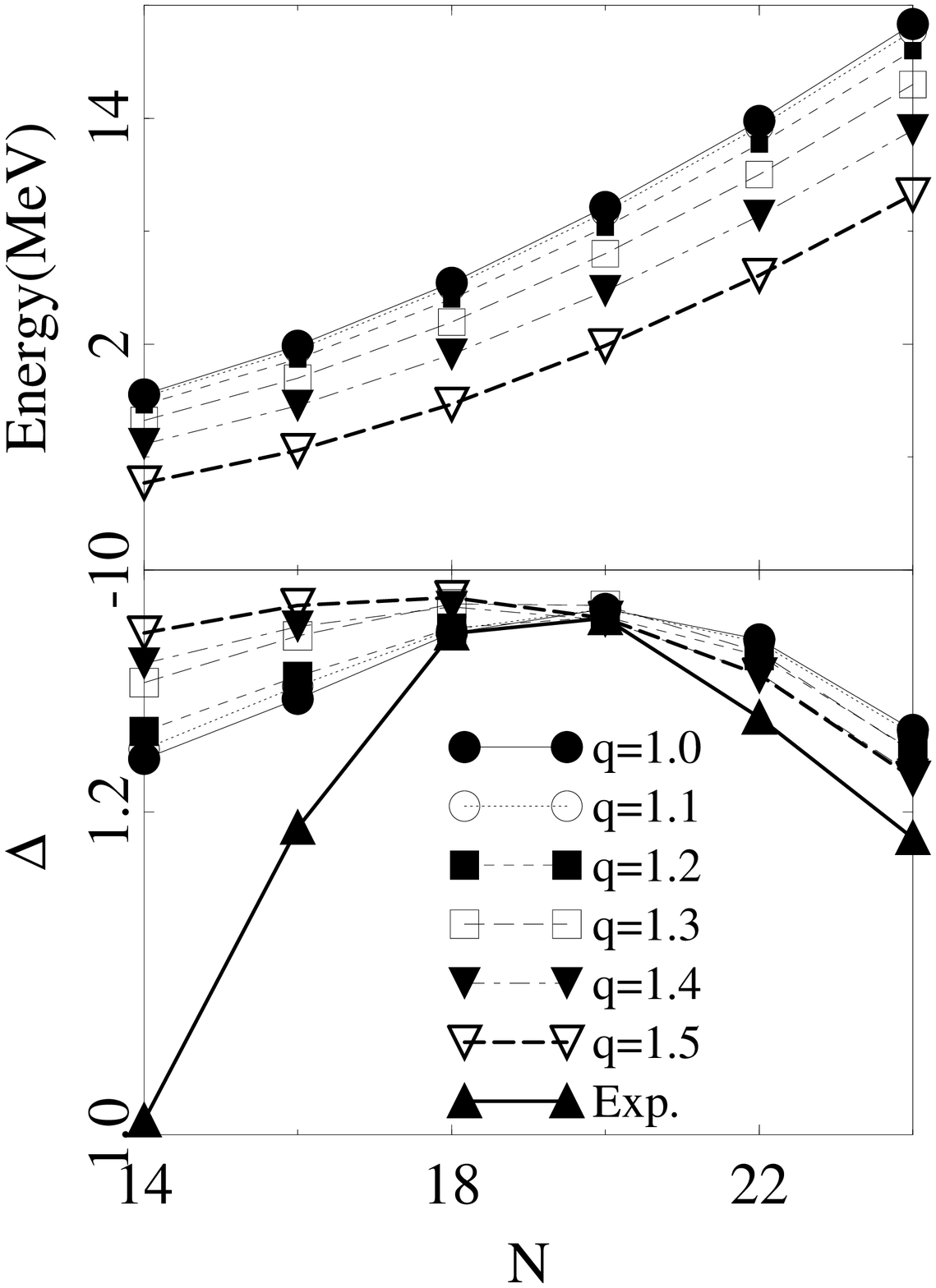,width=6in,height=8in}
\caption{Calculated (a) $\Delta $ vesus N and (b) BCS Energy versus N, for
 $q=1.0,1.1,1.2,1.3,1.4,1.5$ and corresponding $G$ value chosen
to reproduce the empirical neutron gap for $ ^{120}$Sn in each case.}
\label{fig3}
\end{figure}


\begin{references}
\bibitem{BCS57}  J. Bardeen, L. N. Cooper, and J. R. Schrieffer, Phys. Rev. 
{\bf 108}, 1175 (1957) .

\bibitem{she92}  S. Shelly Sharma, Phys. Rev. C{\bf 46,} 904 (1992) .

\bibitem{Jimb85}  M. Jimbo, Lett. Math. Phys. {\bf 10,} 63 (1985); {\bf 11,}
247 (1986) .

\bibitem{Woro87}  Woronowicz, publ. RIMS (Kyoto University) {\bf 23, }117
(1987) ; Commun. Mat. Phys. {\bf 111,} 613 (1987) .

\bibitem{Pasq88}  Pasquier, Nucl. Phys. {\bf B295,} 491 (1988) ; Commun.
Mat. Phys. {\bf 118, }355 (1988) .

\bibitem{Macf89}  Macfarlane A. J., J. Phys. A{\bf 22, }4581 (1989) .

\bibitem{Bied89}  Biedenharn L. C., J. Phys. A{\bf 22, }L873 (1989) .

\bibitem{she94}  S. Shelly Sharma and N. K. Sharma, Phys. Rev. C{\bf 50, }%
2323 (1994) .

\bibitem{Bona92}  D. Bonatsos, J. Phys. {\bf A 25, }L101 (1992) .

\bibitem{Bona94}  D. Bonatsos, C. Daskaloyannis and A. Faessler, J. Phys. 
{\bf A 27, }1299 (1994) .

\bibitem{Bona99}  Dennis Bonatsos and C. Daskaloyannis, Prog. Part. Nucl.
Phys. {\bf 43, }537 (1999) .

\bibitem{law}  R. D. Lawson, $Theory$ $of$ $the$ $Nuclear$ $Shell$ $Model$
(Clarendon, Oxford, 1980)
\end{references}
\end{document}